\documentclass[singlecolumn,aps,prl,preprintnumbers]{revtex4}
\usepackage{bbm}
\usepackage[latin9]{inputenc}
\usepackage{amsmath}
\usepackage{amssymb}
\usepackage{graphicx}
\usepackage{mathrsfs}
\usepackage{amsfonts}
\usepackage{amsthm}
\usepackage{color}
\usepackage{txfonts}
\providecommand{\U}[1]{\protect\rule{.1in}{.1in}}
\makeatletter
\@ifundefined{textcolor}{}
{
\definecolor{BLACK}{gray}{0}
\definecolor{WHITE}{gray}{1}
\definecolor{RED}{rgb}{1,0,0}
\definecolor{GREEN}{rgb}{0,1,0}
\definecolor{BLUE}{rgb}{0,0,1}
\definecolor{CYAN}{cmyk}{1,0,0,0}
\definecolor{MAGENTA}{cmyk}{0,1,0,0}
\definecolor{YELLOW}{cmyk}{0,0,1,0}
}
\makeatother
\begin{document}
\title{Photonic orbital angular momentum transfer and magnetic skyrmion rotation}
\author{Wenrui Yang}
\author{Huanhuan Yang}
\author{Yunshan Cao}
\email[Corresponding author: ]{yunshan.cao@uestc.edu.cn}
\author{Peng Yan}
\email[Corresponding author: ]{yan@uestc.edu.cn}
\affiliation{School of Electronic Science and Engineering and State Key Laboratory of Electronic Thin Film and Integrated Devices, University of
Electronic Science and Technology of China, Chengdu 610054, China}

\begin{abstract}
Magnetic skyrmions are chiral quasiparticles that show promise for future
spintronic applications such as skyrmion racetrack memories and logic devices
because of their topological stability, small size (typically $\sim3-500$ nm), and ultralow threshold force to drive their motion. On the other hand, the ability of light to carry and deliver orbital angular momentum (OAM) in the form of optical vortices has attracted a lot of interest. In this work, we predict a photonic OAM transfer effect, by studying the dynamics of magnetic skyrmions subject to Laguerre-Gaussian optical vortices, which manifests a rotational motion of the skyrmionic quasiparticle around the beam axis. The topological charge of the optical vortex determines both the magnitude and the handedness of the rotation velocity of skyrmions. In our proposal, the twisted light beam acts as an optical tweezer to enable us displacing skyrmions over large-scale defects in magnetic films to avoid being captured.

\end{abstract}
\maketitle
\section{Introduction}

A skyrmion is a swirling noncoplanar texture originally
introduced by Skyrme a half century ago, as a hypothetical particle in the
baryon theory \cite{Skyrme}, and later was observed in many condensed matter
systems, such as liquid crystals, superfluids, and chiral magnets. The magnetic skyrmion, a topologically protected spin texture with a quantized
topological charge
\cite{Nagaosa,Fert,Sampaio,Iwasaki1,Buettner,Romming,Jiang1}, has been a
prominent topic of spintronics since the first experimental observations of
skyrmion lattices in bulk noncentrosymmetric magnets \cite{Muehlbauer,Yu} and
thin films \cite{Heinze}. Depending on the magnetization rotation, three distinct types of skyrmions have been
observed experimentally, referred to as N\'{e}el skyrmion, Bloch skyrmion and antiskyrmion \cite{Parkin}. The manipulation of skyrmions is of great importance
and interest: skyrmions can be driven by spin-polarized current
\cite{Zang,Lin1,Iwasaki2,Huaiyang,Roland}, magnetic field or electric-field gradients
\cite{Everschor,Liu,Wang}, temperature gradients \cite{Kong,Lin2,Mochizuki},
and spin waves \cite{Yan,Iwasaki3,Schuette,Oh}. However, the genuine skyrmion
Hall effects \cite{Jiang2,Litzius} tend to result in skyrmion accumulations at
the edge of devices, although strongly coupled bilayer-skyrmions \cite{Zhang}
were proposed hopefully to overcome the problem. Crystal imperfections
\cite{Mueller,Reichhardt} on the other hand may capture or stop skyrmions.
These difficulties hinder the precise manipulation of the skyrmion motion by
the mentioned control methods. Thus, it should be very interesting and important if one can find other effective control
methods and principles to manipulate skyrmions in magnetic thin films.

Since the theoretical work of Poynting \cite{Poynting} and the experiments by
Beth \cite{Beth1,Beth2}, it has been known that light can carry angular
momentum that is associated with circular polarization and arises from the
spin of individual photons and is termed spin angular momentum. Following the
pioneering work by Allen \emph{et al}. \cite{Allen}, it was realized that
light can also carry orbital angular momentum (OAM) with helical phase fronts
characterized by an $\text{exp}(il\varphi)$ azimuthal phase dependence. Such
twisted lights have a phase dislocation on the axis that is sometimes referred
to as an optical vortex \cite{Basistiy}. The OAM in the light propagation
direction has the discrete value of $l\hbar$ per photon. Integer $l$ is called
the topological charge of the twisted photon. A number of demonstrations and
applications of twisted lights have been brought forward
\cite{Andrews}. They range from optical data storages to
quantum communications, and black holes, among others \cite{Nicolas,Bozinovic,Willner,Tamburini}. Optical vortices
with high OAM have been achieved using spiral phase plates, computer-generated
holograms, mode conversions, spatial light modulators, etc. Very recently, OAM
quantum number up to 10,010 by means of spiral phase mirrors was reported
\cite{Fickler}.

\begin{figure}[ptbh]
\begin{centering}
\includegraphics[width=0.7\textwidth]{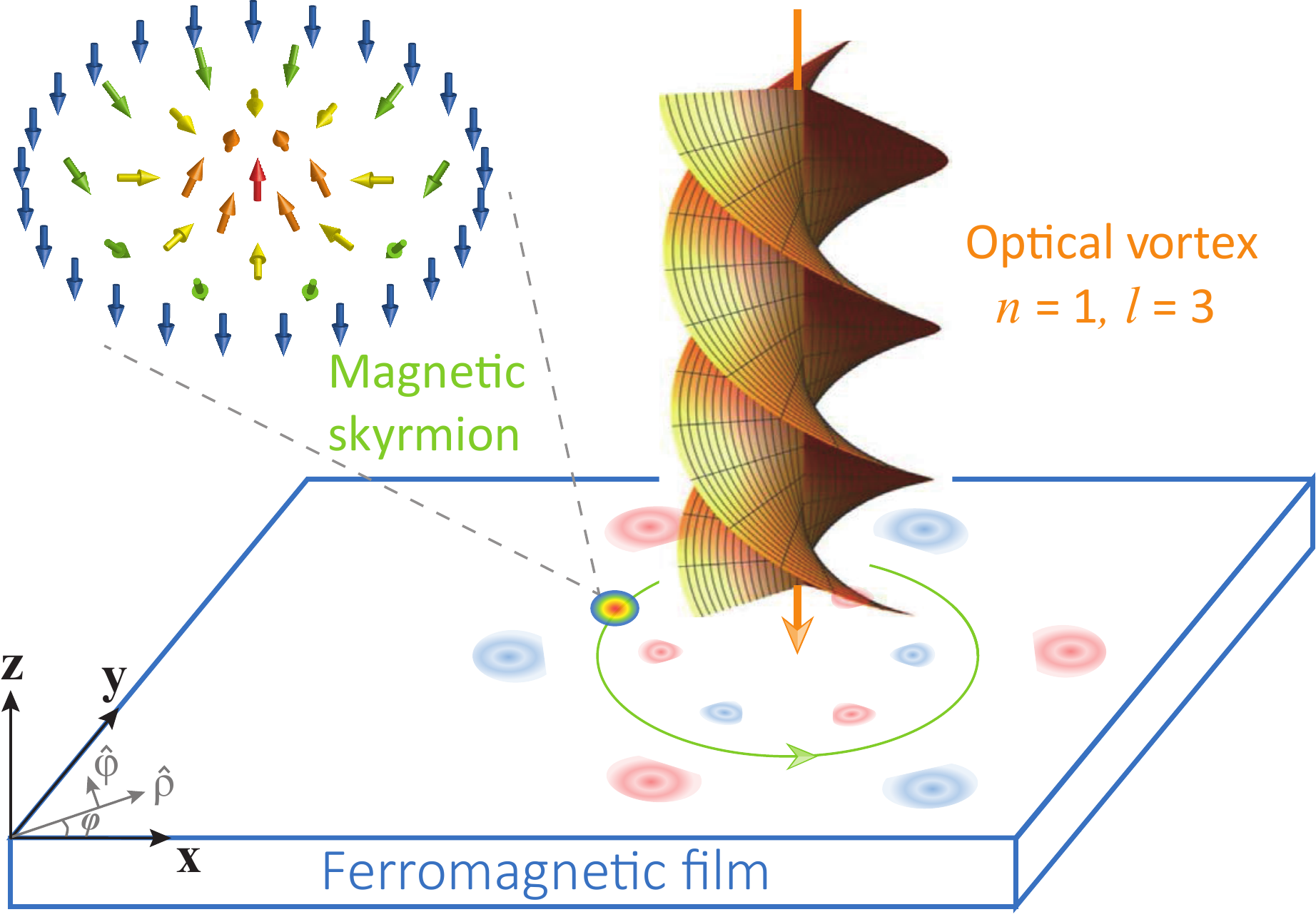}
\par\end{centering}
\caption{Schematic of the rotational motion of a N\'{e}el skyrmion in a thin
ferromagnetic film driven by an optical vortex with radial index $n=1$ and OAM quantum number $l=3$. The solid
circle with a red core represents the skyrmion. The flower-like pattern (pink and blue spots) sketches the
induced magnetization profile by the optical vortex field shinning on the magnetic film. In the main text, the
origin of Cartesian coordinates coincides with the beam center,
while it does not in the figure for clarity.}%
\end{figure}

In this work, we let skyrmions enter the field of OAM. We theoretically
propose to use Laguerre-Gaussian (LG) optical vortices to manipulate the skyrmion
dynamics via the OAM transfer. A rotational motion of an isolated skyrmion is found (see Fig.
1). We show that the topological charge $l$ of the optical vortex plays a key
role in driving the skyrmion motion: a positive $l$ brings about an anticlockwise rotation of
skyrmions around the beam axis, while a negative $l$ results in
a clockwise one. Finally, we demonstrate that optical vortices like tweezers can grip
a skyrmion to overfly large-scale defects in magnetic films to avoid
being captured.

\section{Laguerre-Gaussian beam and OAM transfer}

The twisted light beam with LG mode is a special solution of Maxwell's Eqs. under
the paraxial approximation for electromagnetic waves in vacuum. At the
focal plane $z=0,$ the vortex magnetic field is given by \cite{Allen}
\begin{equation}
\mathbf{B}_{n,l}\left(  \rho,\varphi,t\right)  =B_{0}\frac{\left(  \frac{\rho
}{w}\right)  ^{\left\vert l\right\vert }e^{-\frac{\rho^{2}}{w^{2}}}%
L_{n}^{\left\vert l\right\vert }\left(  \frac{2\rho^{2}}{w^{2}}\right)
}{\sqrt{w}}e^{i\left(  \omega t-l\varphi\right)  }\mathbf{e}_{p}, \label{LG}%
\end{equation}
in the cylindrical coordinate $\left(  \rho,\varphi,z\right)  $ with the
radial coordinate $\rho$, the azimuthal angle $\varphi$, and the coordinate
$z$ along the wave propagation direction (Fig. 1). Here $w$ is the width of
the beam waist, $B_{0}$ is a constant, $L_{n}^{\left\vert l\right\vert }$ is
the generalized Laguerre function, and $\mathbf{e}_{p}$ is the polarization
vector. Two integers $n$ and $l$ denote the radial index and OAM quantum number \cite{OAM},
respectively.
In this work, we focus on linearly polarized lights and assume
$\mathbf{e}_{p}=\mathbf{x}$.

Optical vortex can transfer its OAM to skyrmions via a coherent Zeeman
coupling between photon magnetic fields and local magnetic moments. One thus expect a skyrmion rotation after it absorbs the OAM of twisted lights, besides a possible ultrafast generation of skyrmions \cite{Fujita1,Fujita2}, because a rotating skyrmion manifests itself carrying orbital angular momentum
\cite{Schuette,Tchernyshyov}. In the following, we address this idea both theoretically and numerically.

\section{Theoretical consideration based on Thiele equation}

We consider an isolated N\'{e}el
skyrmion whose spin texture is illustrated in Fig. 1 (Bloch skyrmions and antiskyrmions are also interesting). Magnetization dynamics generally is modeled by the Landau-Lifshitz-Gilbert (LLG) equation \cite{Gilbert}:%
\begin{equation}
\frac{\partial\mathbf{m}}{\partial t}=-\gamma\mathbf{m}\times\mathbf{B}%
_{\text{eff}}+\alpha\mathbf{m}\times\frac{\partial\mathbf{m}}{\partial
t},\label{LLG}%
\end{equation}
where $\mathbf{m}=\mathbf{M}/M_{s}$ is the magnetization unit vector with
saturation magnetization $M_{s},$ $\gamma$ is the gyromagnetic ratio,
$\alpha\ll1$ is the Gilbert damping constant, and $\mathbf{B}_{\text{eff}}$ is the local effective field consisting of isotropic exchange coupling, uniaxial
anisotropy along $z-$axis, the interfacial Dzyaloshinskii-Moriya interaction
favoring N\'{e}el skyrmions, and the optical vortex field (\ref{LG}). Finite temperature effects can be taken into account by including a Gaussian stochastic magnetic field satisfying fluctuation-dissipation theorem \cite{Brown,Yan2}. We focus on skyrmion dynamics at zero temperature, if not stated otherwise. The Thiele approach \cite{Thiele,Oleg} is convenient to derive the analytical
solution, where the magnetization dynamics is encoded in the time
evolution of the skyrmion position and velocity, $\mathbf{m}%
=\mathbf{m}\left(  \mathbf{r}-\mathbf{R}\left(  t\right)  ,\mathbf{V}\left(
t\right)  \right)  $ with $\mathbf{R}\left(  t\right)  $ the
skyrmion center and $\mathbf{V}\left(  t\right)  =\partial_{t}\mathbf{R}\left(
t\right)  $ its velocity. We thus obtain
\begin{equation}
-\mathcal{M}\partial_{tt}\mathbf{R}+\mathbf{G}\times\partial_{t}\mathbf{R}-\alpha\mathcal{D}%
\cdot\partial_{t}\mathbf{R}+\mathbf{F}=0,\label{Thiele}%
\end{equation}
where $\mathcal{M}$ multiplying $M_{s}d/\gamma$ is the effective skyrmion mass
originating from the skyrmion deformation and/or dissipation \cite{Imam,Roberto}, $d$ is the thickness of the film, $\mathbf{G}=-4\pi Q\mathbf{z}$
is the gyromagnetic coupling vector with the skyrmionic topological charge
$Q=\left(  1/4\pi\right)  \int\mathbf{m}\cdot\left(  \partial_{x}%
\mathbf{m}\times\partial_{y}\mathbf{m}\right)  dxdy$, $\mathcal{D}%
_{ij}=\left(  1/4\pi\right)  \int\left(  \partial_{i}\mathbf{m}\cdot
\partial_{j}\mathbf{m}\right)  dxdy$ is the dissipative force tensor, and $\mathbf{F}=-\delta \mathcal {H}/\delta \mathbf{R}$ is the driving force with $H$ the total magnetic energy/Hamiltonian \cite{Slonczewski}. In general, the force $\mathbf{F}$ includes the coupling between the skyrmion with both the optical field and the rotating flower-like magnetization profile (see Fig. 1). Obtaining the analytical expression of the effective force is obviously extremely difficult. As a first approximation, we assume that the skyrmion-flower coupling can be included into the effective skyrmion mass $\mathcal{M}$ and that only the optical field contributes to the effective force. We then arrive at a simple expression $\mathbf{F}=-\int\nabla\mathbf{m}\cdot\mathbf{x}\operatorname{Re}\left(
B_{n,l}\right) dxdy$. The value of $Q$ depends on the detailed spin
profile, e.g. $Q=+1$ for the skyrmion shown in Fig. 1. If $\mathbf{G}\neq0,$
the Lorentz-like force is much larger than the dissipative one. We thus can
ignore the friction in (\ref{Thiele}), and obtain the gyration velocity of the
skyrmion $\mathbf{V}=V\hat{\varphi}$ with $V$ satisfying%
\begin{equation}
\frac{\mathcal{M}V^{2}}{R}+4\pi QV+F_{r}=0,\label{Velocity}%
\end{equation}where $R$ is the gyration radius of skyrmion, i.e., the distance from the
skyrmion to the optical vortex center, and $F_{r}$ is the radial component of the driving
force. Equation (\ref{Velocity}) has two
circular modes with frequencies $\Omega_{\pm}/2\pi=-Q/\mathcal{M}\pm
\sqrt{\left(  Q/\mathcal{M}\right)  ^{2}-F_{r}/\left(  4\pi^{2}\mathcal{M}
R\right)  }$. The friction term in (\ref{Thiele}) then can be treated perturbatively, which leads to a small velocity correction along the radial direction $\delta \mathbf{v}=-\alpha \mathcal{D} V /\left(4\pi Q+\mathcal{M}V/R\right)\hat{\rho}$. In the perturbative calculation, we have assumed a constant $R$. However, the restoring force $\mathbf{F}$ may slow down this radial velocity $\delta \mathbf{v}$ as $R$ varies, and thus leads to a radial oscillation (shown in the numerical simulations below).

We emphasize that the predicted rotational motion is not from the
geometric confinement due to boundary edges \cite{Dai,Kim,Moon} of the film, but from the optical trapping and OAM transfer.
A spatially homogeneous time-oscillating field considered by Moon \emph{et al}. \cite{Moon} can induce an unidirectional skyrmion motion with a helical trajectory, where the skyrmion Hall effect was not avoided.
Here, the skyrmion gyrates locally
around the optical vortex center, instead of either unidirectionally or along device edges,
thus in principle avoiding accumulation or annihilation by boundaries.

\section{Micromagnetic simulations}

To demonstrate the time evolution of the skyrmion motion, we solved numerically the full LLG
Eq. (\ref{LLG}) using the micromagnetic simulation codes MuMax3 \cite{Vansteenkiste,Pizzini,YZhou,XCZhang2}. We used magnetic parameters for Pt/Co/AlO$_{x}$ system with an exchange constant $A=15$ pJ m$^{-1}$, a uniaxial anisotropy $K_{z}=0.8$ MJ
m$^{-3},$ a saturation magnetization $M_{s}=0.58$ MA m$^{-1}$, a
Dzyaloshinskii-Moriya constant $D=3.5$ mJ m$^{-2},$ and a Gilbert damping $\alpha=0.01.$ The skyrmion radius $r_{\text{sk}}$ is about $10$ nm. Materials supporting skyrmions with a larger radius and a stronger thermal stability at room temperature are also interesting. For the majority of results presented in this paper, we consider a
ferromagnetic thin film with length $200$ nm, width $200$ nm and
thickness $1$ nm, which was discretized using 100 $\times$ 100 $\times$ 1
finite difference cells. The spin Hamiltonian in the lattice is then $\mathcal {H}=-A'\Sigma_{<i,j>}\mathbf{m}_{i}\cdot\mathbf{m}_{j}-D'\Sigma_{<i,j>}(\mathbf{m}_{i}\times\mathbf{m}_{j})\cdot\mathbf{z}-K'_{z}\Sigma_{i}(\mathbf{m}_{i}\cdot\mathbf{z})^{2}-M_{s}\Sigma_{i}\mathbf{B}_{n,l,i}\cdot\mathbf{m}_{i}$, where $\mathbf{m}_{i}$ is the unit magnetization vector at $i$ site, $<i,j>$ runs over the nearest neighborings, with parameters $A'=187.5$ meV, $D'=87.5$ meV, and $K'_{z}=40$ meV in lattices. In the simulations, we consider simple optical vortices with $n=1$. Each $B$ field has one zero at $\rho=\rho_{0}$ (besides $\rho=0$) and
two local extremums at $\rho=\rho_{1}$ and $\rho_{2}$ $\left( 0<\rho_{1}%
<\rho_{0}<\rho_{2}\right) $. The field gradient at $\rho=\rho_{0}$ increases
sharply with increasing index $l$ (see Appendix for Fig. 5). In order to reduce the skyrmion deformation caused by the twisted light, we place the skyrmion initially at the zero point $\rho=\rho_{0}$ of the optical field. We note that, in general, the magnetic/optical constraint works very well when we initially put the skyrmion in the region $\left( 0<\rho_{1}<\rho<\rho_{2}\right)$. We set $w=10$ nm (cf. wavelength $\lambda\sim$ 10 cm for GHz microwaves in vacuum). We should point out that the width of the
light-beam waist $w$ usually cannot be smaller than a half of the wavelength
of light due to the diffraction limit. However, thanks to the plasmonics
techniques \cite{Gorodetski,Barnes}, it is possible to achieve focused twisted
light with beam width much smaller than the wavelength \cite{Heeres,Cui}. On the other hand, although it is computationally too expensive to simulate the gyration of nano-scale skyrmions driven by centimeter-scale microwave vortices if the diffraction barrier is respected, the present results can transfer to those large scales by a proper scaling of the topological charge $l$ with respect to the optical vortex size $w$ based on an angular momentum conservation argument: The dissipation rate of the OAM of a rotating skyrmion is $\alpha\mathcal{D}\mathbf{R}\times\partial_{t}\mathbf{R}$ which should be compensated by optical vortices with OAM $l\hbar$ per photon [This can be proved by taking a cross product with $\mathbf{R}$ at both sides of Eq. (\ref{Thiele})]. So, one should keep $l\propto R$ to maintain comparable skyrmion rotation velocities when a much wider beam ($w\sim R$) is used. Numerical results below justifies this argument.

\begin{figure*}[ptbh]
\begin{centering}
\includegraphics[width=\textwidth]{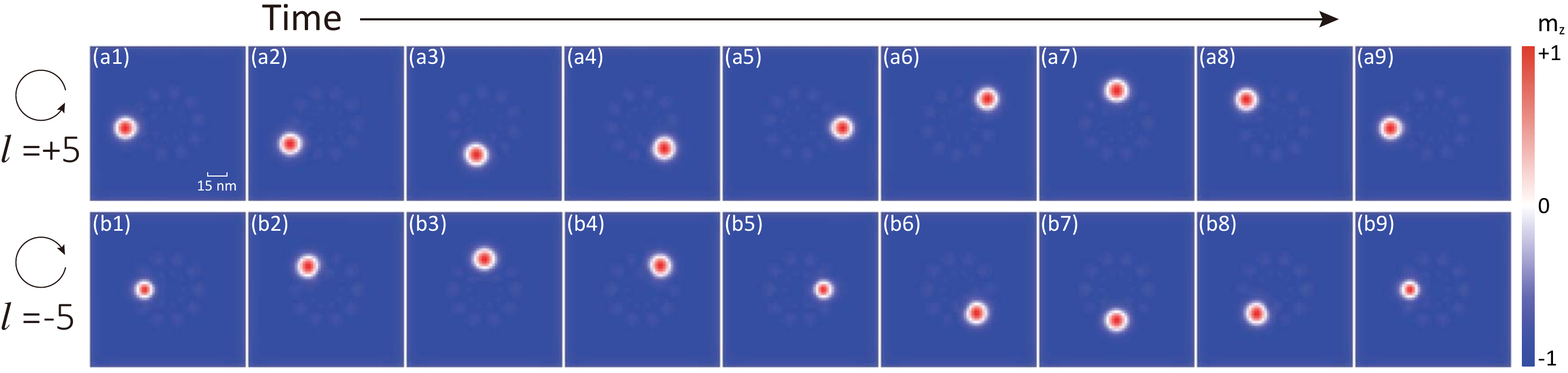}
\par\end{centering}
\caption{Time evolution of an isolated N\'{e}el skyrmion under optical
vortices with OAM quantum number $l=+5$ (a1)-(a9) and $l=-5$ (b1)-(b9). The time intervals
between successive snapshots in two cases are $0.8$ ns and $0.5$ ns,
respectively. A positive (negative) topological charge induces an
anticlockwise (clockwise) rotation of a skyrmion around the beam axis.}%
\end{figure*}

In Figs. 2(a1)-(a9), we show an anticlockwise rotation of the skyrmion about
the beam center due to an application of an optical vortex with OAM quantum number $l=+5$,
frequency $f=\omega/2\pi=1$ GHz, and coefficient $B_{0}=10^{-4}$ T m$^{1/2}$ corresponding to
a peak magnetic field 1.7 T of the optical vortex. The initial distance $\rho_{0}$ between the skyrmion
and the optical vortex center is about $20$ nm. We find quite a stable
skyrmion rotation with negligibly small fluctuations in both
the radius and shape (see \textcolor{blue}{Visualization 1}) indicating a nearly massless skyrmion. The period of
rotation is about $6$ ns. The linear
velocity of the skyrmion $V_{+}$ is close to $-34.6$ m s$^{-1}$ (positive velocity is defined for clockwise rotation). The velocity is comparable to that achieved by spin-polarized currents with electrical
current density of $10^{10}$ A m$^{-2}$ \cite{Fert}. The radius of gyration $R_{+}\approx 33$ nm is larger
than $\rho_{0}$, because the viscous drag $\delta \mathbf{v}$ is outward in the presence of a negative gyration velocity. This stable skyrmion rotation can survive at finite temperatures up to 200 K, without obvious velocity reduction (see Appendix for finite temperature simulations and Fig. 6).

To prove that the rotational motion is indeed due to the OAM transfer from optical vortices to the skyrmion, we provide two more numerical evidences: (i) we reverse the
helicity of the optical vortex from $l=+5$ to $l=-5$ without changing the rest
parameters. Figures 2(b1)-(b9) demonstrate a striking reversal of the
rotation direction of skyrmions, as expected. The rotation
period becomes $4$ ns. The linear velocity $V_{-}$ of the skyrmion is estimated to be $31$ m s$^{-1}$, with a compressed radius of gyration $R_{-}\approx19.7$ nm slightly smaller than $\rho_{0}$ since the
viscous drag $\delta \mathbf{v}$ now becomes inward. An interesting skyrmion
breathing, i.e., skyrmion radius regularly expands and contracts, is found. The observed asymmetry between $|V_{+}|$ and $|V_{-}|$ can be well explained by the effective skyrmion mass. From Eq.
(\ref{Velocity}), we derive the emerging skyrmion mass by
equating $\mathcal{M}V_{-}^{2}/{R_{-}}$ with $4\pi Q (-V_{+}-V_{-})$, assuming $F_{r}(l=5)\approx-F_{r}(l=-5)$. After
including the coefficient $M_{s}d/\gamma=3.3\times10^{-15}$ J s
m$^{-2}$, we evaluate the skyrmion mass $\mathcal{M}\approx3.06\times10^{-24}$
kg. To demonstrate the robustness of the optical confinement, we have tested the case that the skyrmion carries a finite initial velocity achieved by applying an in-plane spin polarized electric current. We find that a small initial velocity, e.g., $16.7$ m/s (driven by a current with density $2\times10^{11}$ A/m$^{2}$), cannot destroy the magnetic constraint. To test this issue further, we increase the current density up to $2\times10^{13}$ A/m$^{2}$, and still find a very nice magnetic constraint. This can be related to the fact that the skyrmion mass is so small that its inertial effect does not play a key role to overcome the optical trapping. (ii) There still exists a loophole in the above analysis that the skyrmion is probably just interacting with the moving potential minima (the rotating flower) created by the spatially inhomogeneous field. In order to close this loophole, we cut the film into a very narrow annular plate to exclude the effect from the rotating magnetization flower. We observe a fast skyrmion rotation in the confined annular geometry (see \textcolor{blue}{Visualization 2} and the Appendix for Fig. 7). This is a direct evidence to justify our understanding. We also note that switching $Q$ from
$+1$ to $-1$, i.e., spins in the skyrmion core pointing down surrounded by
background spins pointing up, does not reverse the handedness of skyrmion rotation. This can be understood that both the driving force $F$
and skyrmionic topological charge $Q$ are odd
functions of the magnetization $\mathbf{m}$, which thus remains the handedness of the rotation identical for opposite $Q$, cf. Eq. (\ref{Velocity}). This unique feature has motivated us to apply the present idea to
skyrmions in antiferromagnets \cite{Joe,Zhou} which usually operate in terahertz (THz) ranges, because an antiferromagnetic skyrmion can be regarded as a superposition of two ferromagnetic skyrmions with antiparallel spins and opposite topological charges. We indeed numerically observed a much faster antiferromagnetic skyrmion rotation with an extremely high velocity $\sim 1000 $ m s$^{-1}$ driven by THz optical vortices (see \textcolor{blue}{Visualization 3} and the Appendix for the parameters used in the simulation).

\begin{figure}[ptbh]
\begin{centering}
\includegraphics[width=0.9\textwidth]{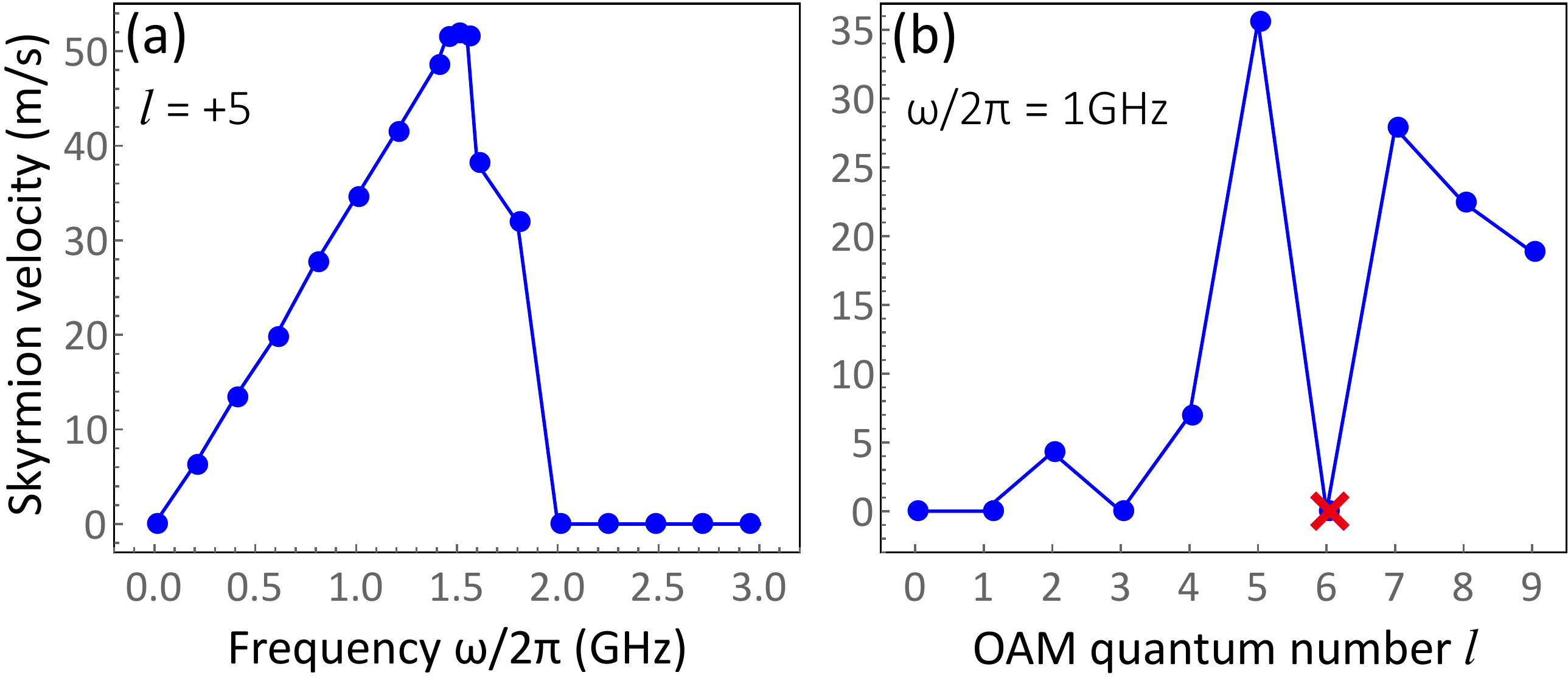}
\par\end{centering}
\caption{Skyrmion velocity as a function of the driving frequency (a) and the
topological charge (b) of optical vortices. The red cross stamps a skyrmion
annihilation. The minus sign of the velocity is dropped.}%
\end{figure}

The rotation velocity of skyrmions as a function of the driving frequency
$\omega$ is shown in Fig. 3(a). The velocity linearly increases with the frequency
below 1.5 GHz and sharply decreases then. The peak is due to the
resonance between the skyrmion circular frequency $\Omega_{+}$ and the vortex frequency
$\omega$. Higher frequency indicates more diluted photon number density $\propto B_{0}^2/\omega$, thus reducing the OAM transfer and lowering the skyrmion velocity drastically. Figure 3(b) shows the OAM dependence of the skyrmion velocity, in
which we fix the frequency at 1 GHz and choose a coefficient $B_{0}%
=6\times10^{-5}$ T m$^{1/2}$ \cite{Max}. The skyrmion velocity is
shown to be sensitive to $l$. At
$l=+6$, the breathing skyrmion moves outward due to the radial drag and breaks down when approaching the position $\rho=\rho_{2}$ where the magnetic field of the optical vortex reaches its local extremum which is so strong to annihilate skyrmions (see \textcolor{blue}{Visualization 4}). However, this annihilation does not happen for smaller $l\left(<6\right)$ because the magnitude of the vortex field is too weak. Larger $l\left(>6\right)$ sustains a stable skyrmion rotation because skyrmions are pulled back by strong restoring forces (radial field gradients) before they hit the local maximum at $\rho=\rho_{2}$. A generic criteria to this instability is that the discriminant $\Delta =\left(  2\pi Q\right)^{2}-F_{r}\mathcal{M}/R$ becomes negative which allows no real solutions of Eq. (\ref{Velocity}). For $l>6$, the skyrmion annihilation does not happen because the corresponding $F_{r}$ is actually much smaller than the one at $l=+6$. This interpretation is based on analyzing the profile of the optical vortex field $B_{n,l}\left(\rho\right)$ plotted in Fig. 5, which has been discussed at the beginning of this session. In our model, the coupling strength between skyrmions and the magnetic profile of the light solely depends on the Zeeman interaction. Since the optical vortex field is non-uniformly distributed in space, the coupling strength varies for different skyrmion position with respect to the beam axis, and for different profile of optical fields as well. The skyrmion velocity dependence on $l$ calculated above indeed changes quantitatively if $B_{0}$ (amplitude of light) and $w$ (waist width) are different, while the driving mechanism is essentially the same. A thorough investigation on skyrmion annihilations process will be presented elsewhere.

\begin{figure}[ptbh]
\begin{centering}
\includegraphics[width=0.5\textwidth]{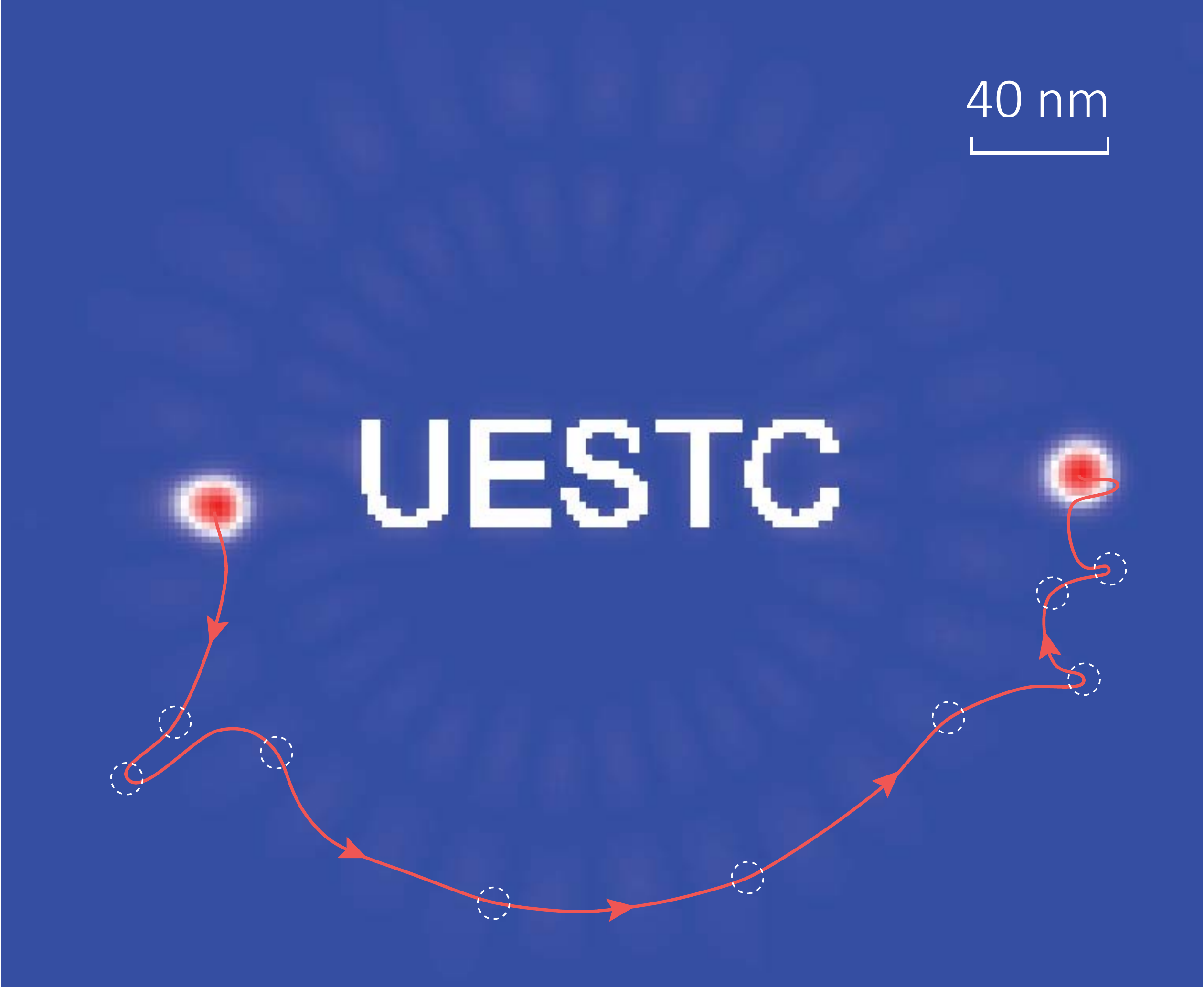}
\par\end{centering}
\caption{Skyrmion overflies patterned antidots. The red curve represents the trajectory of the skyrmion with arrows
indicating its moving direction. Small white dashed circles are snapshots of the
skyrmion. It takes 11.5 ns to deliver the skyrmion from left to the right. }%
\end{figure}

Another merit of our proposal is that a proper design of the gyration orbit can enable
skyrmions to bypass defects in magnetic films. To this end, we numerically
show how optical vortices can displace magnetic skyrmions across a long
distance to overfly patterned antidots (no materials) forming a picture UESTC (white holes in blue magnetic region shown in Fig. 4) on the film of size 400 nm $\times$ 400 nm $\times$ 1
nm. Here UESTC stands for the abbreviation of the institute of the authors (It can be replaced by other patterns). We consider a beam with a larger width $w=35$ nm and an enhanced OAM quantum number $l=+15$. A successful flight for skyrmions without being captured by the
holes is observed, despite of a somewhat lingering process due to the competition between the viscous drag induced radial oscillation and the gyration motion (see \textcolor{blue}{Visualization 5}). The average rotation velocity of the skyrmion is $50$ m s$^{-1}$, which quantitatively supports the earlier argument on how to transfer our results to a larger length scale. If the defects/barriers are placed in the light-induced magnetic profile, rather than the beam center, one may expect that the field strength of the optical vortex should be large enough to help the skyrmion bypass them. Numerical simulations indeed demonstrate this point (not shown).

\section{Conclusion}
In summary, we proposed an all-photonic orbital angular momentum transfer mechanism to manipulate magnetic
skyrmions. This OAM transfer can effectively
drive the skyrmion rotation around the beam axis. Optical vortices like tweezers are
capable of gripping the skyrmion to overfly large-scale barriers without being trapped. The mechanism applies not only to ferromagnetic skyrmions but also to antiferromagnetic ones. Our proposal opens the door for all-optical manipulations of magnetic skyrmions by harvesting the OAM of twisted lights and raises the
challenge to generate micron-/submicron-focused optical vortices. Other types of vortex beams such as electron vortices \cite{Beche,Mafakheri} and their acoustic counterparts \cite{Hong,Diego} without breaking the diffraction limit are also promising candidates to drive skyrmion rotations via magnetoelectric and magnetoelastic couplings, respectively. These should be interesting subjects for future research.

\section{Appendix}
This appendix consists of four parts: The first one shows the spatial distribution of the optical vortex field, the second one details the description of 5 visualizations, the third one calculates the finite temperature effect on the optical vortex driven skyrmion motion, and the last one discusses the skyrmion motion in a narrow annulus.

\subsection{Optical vortex fields}

\begin{figure}[ptbh]
\begin{centering}
\includegraphics[width=0.8\textwidth]{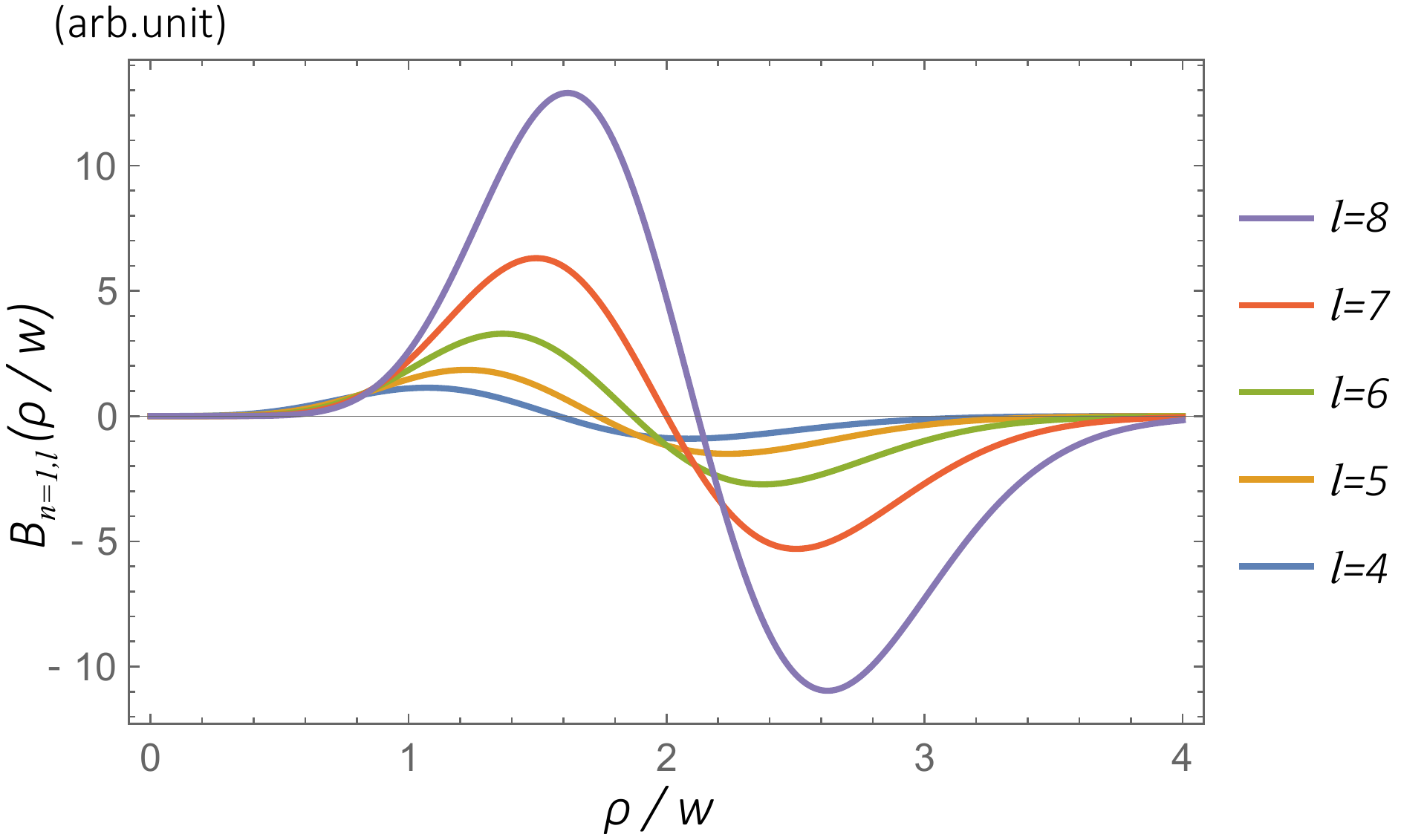}
\par\end{centering}
\caption{Spatial profile of the optical vortex field $B_{n,l}(\rho)$ with $n=1$ for
different index $l$.}
\end{figure}As shown in Fig. 5, each $B$ field has one zero at $\rho=\rho_{0}$ (besides $\rho=0$) and
two local extremums at $\rho=\rho_{1}$ and $\rho_{2}$ $\left( 0<\rho_{1}%
<\rho_{0}<\rho_{2}\right) $. The field gradient at $\rho=\rho_{0}$ increases
sharply with increasing index $l$.

\subsection{Visualizations}
We describe the five visualizations in detail below.

\textcolor{blue}{Visualization 1}: Ferromagnetic skyrmion rotation driven by the optical vortex with OAM quantum number $l=+5$ under
frequency $f=\omega/2\pi=1$ GHz and coefficient $B_{0}=10^{-4}$ T
$\text{m}^{1/2}$. The beam waist width is $w=$10 nm.

\textcolor{blue}{Visualization 2}: Optical vortex driven skyrmion motion in a narrow annular plate (shown in Fig. 7).

\textcolor{blue}{Visualization 3}: To perform simulations of optical vortex driven antiferromagnetic skyrmion motion, we use the material parameters for antiferromagnet KMnF$_3$ (see Ref. [67] in the main text), with the parameters as follows: lattice constant $a=0.5$ nm, saturation magnetization $M_{s}=0.376$ MA m$^{-1}$, antiferromagnetic exchange constant $A=-6.59$ pJ m$^{-1}$, uniaxial anisotropy $K_{z}=0.116$ MJ m$^{-3},$ Dzyaloshinskii-Moriya constant $D=0.7$ mJ m$^{-2},$ and a Gilbert damping $\alpha=0.01$. We consider a film of size 150 nm $\times$ 150 nm $\times$ 0.5 nm. The mesh size for the simulations is 0.5 nm. The applied THz optical vortex carries OAM quantum number $l=-10$, frequency $f=2$ THz, waist width $w=10$ nm, and $B_{0}=1\times10^{-5}$ T m$^{1/2}$. In the simulation, we set a fixed time step $\Delta t=$ 10 fs.

\textcolor{blue}{Visualization 4}: Ferromagnetic skyrmion annihilation process by an optical vortex with OAM quantum number $l=+4$ and $B_{0}=7\times10^{-5}$ T m$^{1/2}$ in a larger ferromagnetic film of size 400 nm $\times$ 400 nm $\times$ 1 nm.

\textcolor{blue}{Visualization 5}: Optical vortices displace ferromagnetic skyrmions to overfly patterned antidots (no materials) UESTC (white holes in blue magnetic region) on the two-dimensional ferromagnetic film of size 400 nm $\times$ 400 nm $\times$ 1 nm. The applied optical vortex carries OAM quantum number $l=+15$ with frequency $f=1$ GHz, width $w=35$ nm, and coefficient $B_{0}=3.3\times10^{-8}$ T m$^{1/2}$. The maximal magnetic field of the optical vortex is 1.29 T. Center of the beam coincides with that of the patterned antidots.

\subsection{Finite temperature calculations}

\begin{figure*}[ptbh]
\begin{centering}
\includegraphics[width=\textwidth]{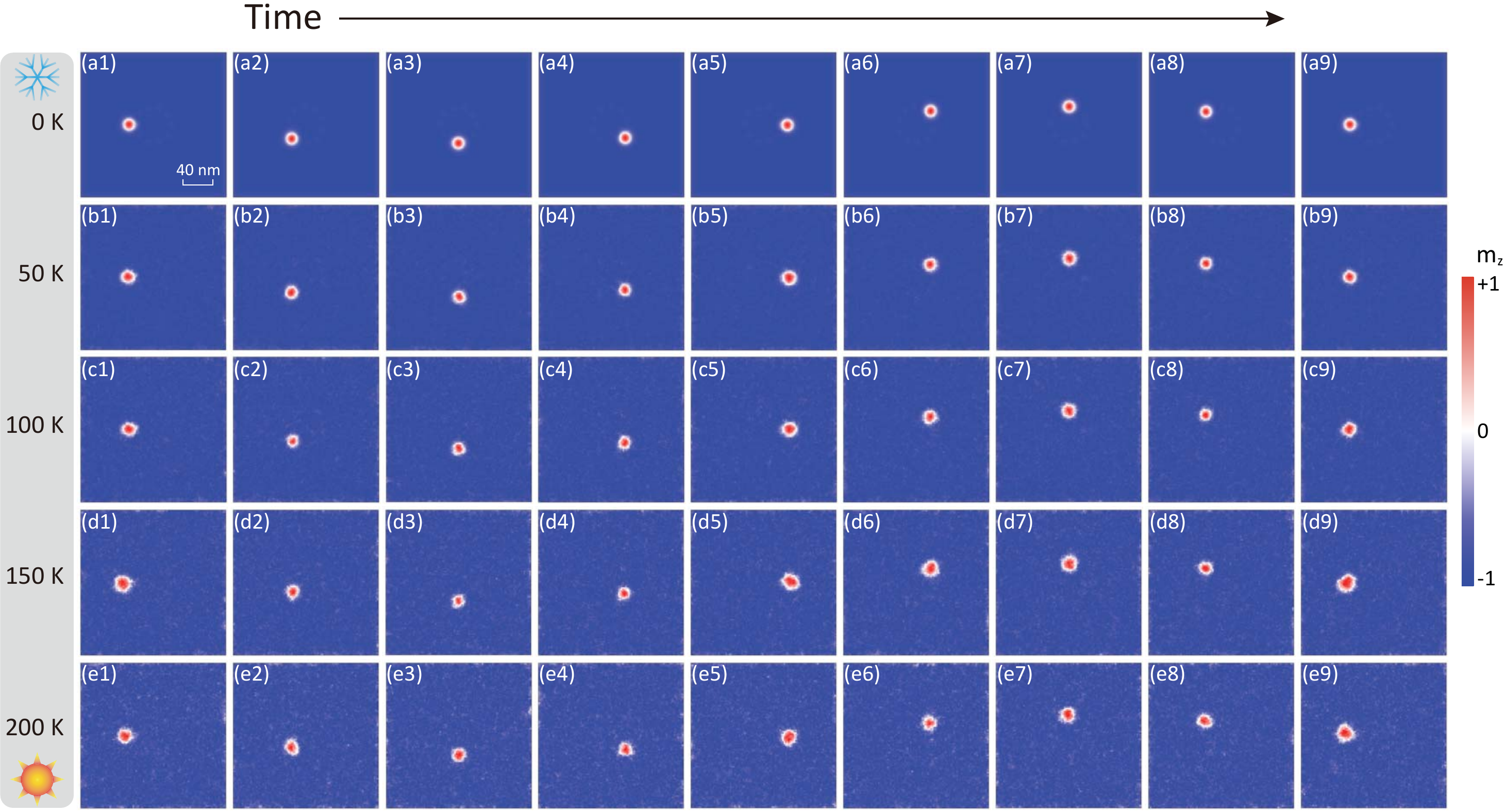}
\par\end{centering}
\begin{flushleft}
\caption{Time evolution of a N\'{e}el skyrmion driven by an optical vortex with
$l$=+5 at different temperatures. The time interval between successive
snapshots for each temperature is $0.8$ ns.}
\end{flushleft}
\end{figure*}

All numerical data presented in the main text are calculated at zero temperature. Here, we take finite temperatures into account by introducing a Gaussian stochastic
magnetic field $\mathbf{h}$ satisfying fluctuation-dissipation theorem
\begin{equation}
\langle h_{i}(\mathbf{r},t)h_{j}(\mathbf{r}^{\prime},t^{\prime})\rangle
=\frac{2\alpha k_{B}T}{\gamma M_{s} \Delta V}\delta(\mathbf{r}-\mathbf{r}^{\prime
})\delta_{ij}\delta(t-t^{\prime})
\end{equation}
into the effective field in Eq. (2) in the main text, with $k_{B}$ the
Boltzmann constant, $T$ the absolute temperature
of thermal bath, and $\Delta V$ the volume of the cell.

Without optical vortices, a single skyrmion can survive in the ferromagnetic
thin film at temperatures up to $T=250$ K. In the presence of an
optical vortex with OAM quantum number $l=+5$, waist width $w=10$ nm, frequency
$f=\omega/2\pi=1$ GHz, and coefficient $B_{0}=10^{-4}$ T m$^{1/2}$, we find quite a
stable skyrmion rotation without obvious reduction of the velocity by
increasing the temperature till $T=200$ K, above which the skyrmion becomes
unstable and can be annihilated by strong thermal fluctuations. This proves
that the optical OAM transfer mechanism is robust against thermal
fluctuations. Figure 6 shows the calculated time evolution of an
isolated N\'{e}el skyrmion under the optical vortex described above, at five
different temperatures: $T=0$ K in Fig. 6(a1)-6(a9), $T=50$ K in Fig. 6(b1)-6(b9), $T=100$ K
in Fig. 6(c1)-6(c9), $T=150$ K in Fig. 6(d1)-6(d9), and $T=200$ K in Fig. 6(e1)-6(e9).

\subsection{Skyrmion rotation in an annular plate}

\begin{figure*}[ptbh]
\begin{centering}
\includegraphics[width=0.8\textwidth]{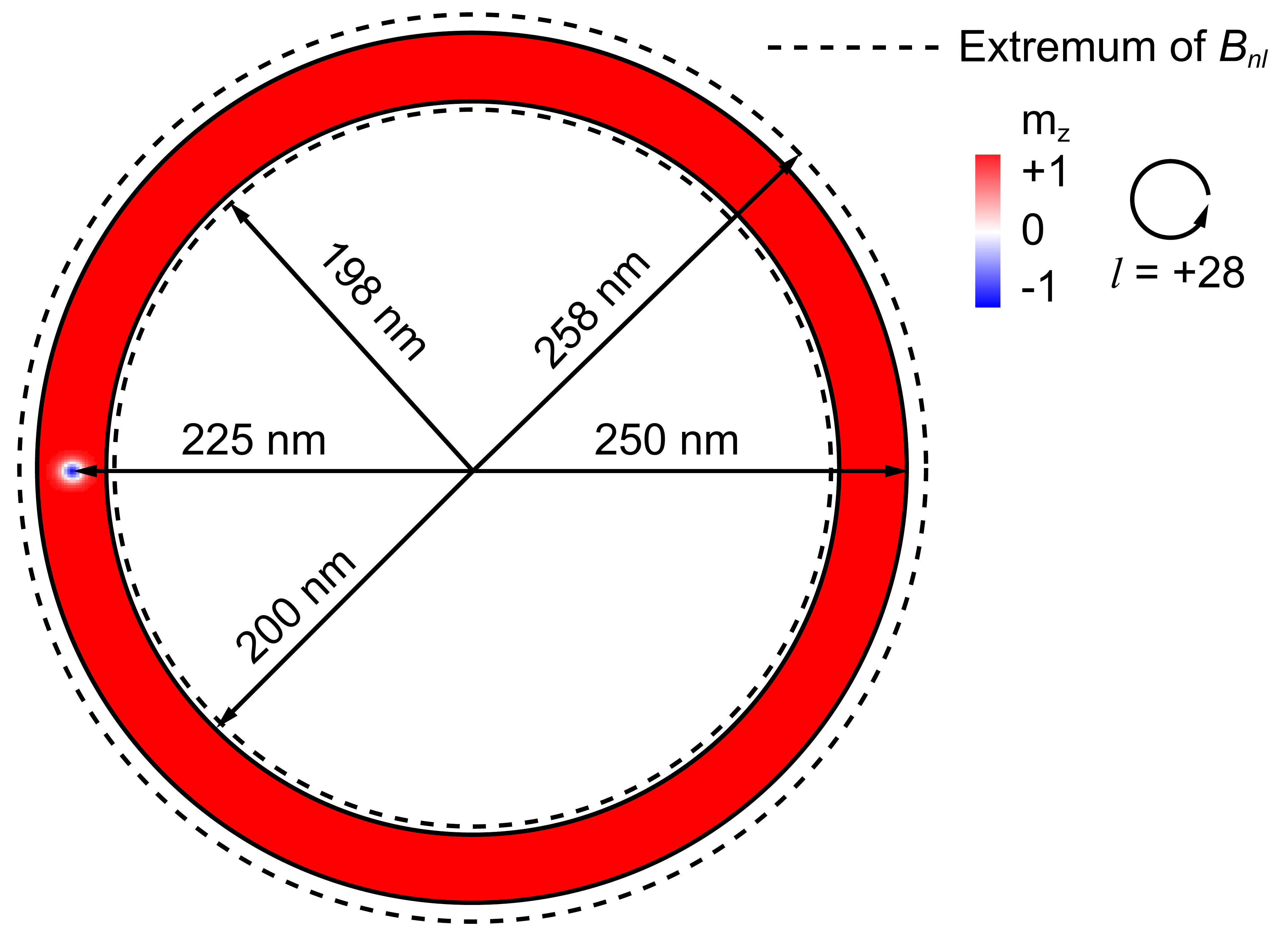}
\par\end{centering}
\caption{Geometry of an annular plate confining the skyrmion driven by an optical vortex with $l=+28$.}
\end{figure*}
The ferromagnetic thin film is cut into a very narrow annular plate with inner radius 200 nm and outer radius 250 nm (shown in Fig. 7). The two dashed circles denote the local extremums of the optical vortex with radius $\rho_{1}=198$ nm and $\rho_{2}=$258 nm, respectively. In this annular geometry, we are able to exclude the effect from the rotating flower-like magnetization profile (or the moving potential minimum) on the skyrmion motion. In the numerical simulation, we utilize an optical vortex with frequency $f=1$ GHz, field coefficient $B_{0}=4.0\times10^{-15}$ T m$^{1/2}$, waist width $w=60$ nm, $n=1$, and OAM quantum number $l=+28$. We use a mesh size 1 nm $\times$ 1 nm $\times$ 1nm. A fixed/pinned boundary condition is adopted for spins at both the inner and the outer edges. The observed period of skyrmion rotation is about 29 ns with an average velocity 48.7 m s$^{-1}$ along the orbit of radius 225 nm (see \textcolor{blue}{Visualization 2}). We thus provide a strong and direct evidence to support our claim of all-photonic orbital angular momentum transfer effect.

\section*{Funding}
National Natural Science Foundation of China (Grants No. 11604041 and 11704060); the
National Key Research Development Program under Contract No. 2016YFA0300801; the National Thousand-Young-Talent Program of China.

\section*{Acknowledgments}
We thank H.Y. Yuan and Z.Y. Wang for useful discussions.


\begin{thebibliography}{99}
\bibitem{OAM}The cycle-averaged angular momentum density
$\mathbf{j}$ associated with the transverse EM field is shown to be
$\mathbf{j}=\varepsilon_{0}\mathbf{r}\times\left\langle \mathbf{E}%
\times\mathbf{B}\right\rangle $, while the total angular momentum of the field
is $\mathbf{J}=\varepsilon_{0}\int\mathbf{r}\times\left\langle \mathbf{E}%
\times\mathbf{B}\right\rangle d\mathbf{r}.$ Here $\varepsilon_{0}$ is the
vacuum permittivity and $\mathbf{E}$ is the electric field. For linearly
polarized LG modes (\ref{LG}), the OAM density in the direction of propagation
is given by $j_{z}=\varepsilon_{0}\omega l\left\vert B\left(  \mathbf{r}%
\right)  \right\vert ^{2}.$ The OAM integrated over the beam is shown to be
equivalent to $l\hbar$ per photon: $J_{z}=l\hbar$.
\bibitem{Max} The corresponding maximal magnetic field of the optical
vortex is 0.56 T, 0.34 T, 0.35 T, 0.44 T, 0.63 T, 1.06 T, 1.87 T, 3.38 T, 7.61
T, and 16.76 T for $l=0,1,2,\cdots,$ and 9, respectively.
\end{thebibliography}
\end{document}